\documentclass[
    ,final            
  ]
  {aipproc}

\layoutstyle{6x9}

\begin{document}

\title{Modelling lightcurves and spectra of transient Anomalous X-ray 
Pulsars}

\classification{97.60.Jd, 98.70.Qy, 98.70.Rz}
\keywords      {radiation mechanisms: non-thermal  --- sources 
(individual):
XTE J1810-197, CXOU J164710.2-455216 ---
stars: magnetic fields --- stars: neutron}

\author{S. Zane}{
  address={MSSL, University
College London, Holmbury St. Mary, Dorking, Surrey, RH5 6NT, UK}
}

\author{A. Albano}{
  address={Department of Physics,
University of Padova, Via Marzolo 8, I-35131 Padova, Italy}
}

\author{R. Turolla}{
  address={Department of Physics,
University of Padova, Via Marzolo 8, I-35131 Padova, Italy}
,altaddress={MSSL, University
College London, Holmbury St. Mary, Dorking, Surrey, RH5 6NT, UK}
}

\author{G.L. Israel}{
  address={INAF-Astronomical Observatory of Rome, via
Frascati 33, I-00040, Monte Porzio Catone, Italy}
}

\author{L. Nobili}{
  address={Department of Physics,
University of Padova, Via Marzolo 8, I-35131 Padova, Italy}
}

\author{L. Stella}{
  address={INAF-Astronomical Observatory of Rome, via
Frascati 33, I-00040, Monte Porzio Catone, Italy}
}

\begin{abstract}
We present the first detailed
{\it joint} modelling of both the timing and spectral properties during 
the
outburst decay of transient anomalous X-ray pulsars. We consider the two 
sources  XTE J1810-197 
and CXOU J164710.2-455216, and describe the source decline in
the framework of a twisted magnetosphere model, using Monte Carlo
simulations of magnetospheric scattering and mimicking localized heat
deposition at the NS surface following the activity. Our results support a 
picture in which a
limited portion of the star surface close to one of the magnetic   
poles is heated at the outburst onset. The subsequent evolution is
driven both by the cooling/varying size of the heated cap and by a
progressive untwisting of the magnetosphere.
\end{abstract}

\maketitle


\section{Introduction}

Anomalous X-ray pulsars (AXPs) and soft gamma repeaters (SGRs) are two
classes of X-ray sources whose phenomenology may be explained by the
presence of a magnetar\footnote{See this volume for alternative 
scenarios}, i.e. an ultra-magnetized neutron star (NS) with a magnetic 
field as high
as $B \sim 10^{14}-10^{15}$~G. Both AXPs and SGRs exhibit variability in 
their persistent X-ray emission over a timescale of years, characterized 
by large flux changes (factor 10-100) and drastic alterations in the 
spectral and timing properties. The recently discovered Transient AXPs 
(TAXPs) represent the far end of this behaviour. Both XTE 1810-197 and 
CXOU J164710.2-455216 underwent a sudden outburst during which their X-ray flux
increased by more than two orders of magnitude and the emission of
SGR-like bursts was observed. Their post-outburst decline has continued 
for a few years after the event and has been monitored extensively with 
{\it XMM-Newton}. 

These observations provided a large sample of high-quality data, the
analysis of which offers an unprecedented opportunity to unveil the
mechanisms responsible for SGRs/AXPs activity. Here we present a 
detailed
{\it joint} modelling of both the timing and spectral properties during 
the
outburst decay of these two TAXPs. We describe the source decline in
the framework of a twisted magnetosphere model, and make use of the 
spectral models presented and applied by \cite{ntz} (NTZ in the 
following), and \cite{za09}, who studied radiative transfer in a globally twisted magnetosphere by means of 
a 3D Monte Carlo code. Each model is characterized by the magnetospheric 
twist angle $\Delta\phi$, the electron (constant) bulk velocity $\beta$,
and the seed photon temperature $kT$. The polar field is fixed at
$B=10^{14} \ {\rm G}$. 
Besides, in order to mimick localized heat deposition at the NS surface 
following the activity, we generalize the original application 
by NTZ by allowing for a complex thermal surface map. For details on 
the method of analysis and the observational data, we refer to the 
original paper by \cite{ale} and references therein.

\section{TAXP Analysis}

XTE 1810-197 was serendipitously discovered in 2003 with {\it RXTE}, and a 
comparison with archival {\it RXTE} data showed it produced an outburst 
around Nov 2002, followed by a monotonic decline of the X-ray flux 
\cite{ib04, is04, gh05, gh07, pe08, be09}. 
In this investigation we use 8 {\it XMM-Newton} observations of the 
source, covering a baseline of 4 years between Sep 2003 and Sep 2007. 
As for CXOU J164710.2-455216, we use 6 {\it XMM-Newton} observations 
made from Sep 2006 to Aug 2009, across the Nov 2006 outburst of the 
source \cite{muno07, is07}. 

\begin{figure}
\includegraphics[width=1.6in,angle=0]{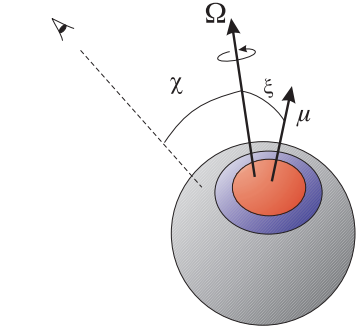}
\caption{A schematic view of the NS.  $\xi$, $\chi$ are the viewing 
angles, 
$\Omega$ and
$\mu$ are the star spin and magnetic axis,  
the dashed line is the line-of-sight. The star surface is 
divided into three regions: a hot polar cap (red), a warm corona (blue)
and a colder zone (grey). Reproduced by permission of the AAS from 
\cite{ale}.\label{fig1}}
\end{figure}

Our strategy proceeds as follows. We first jointly fit the pulse profiles
in three energy bands (0.5-10 keV, 0.5-2 keV, 2-10 keV), allowing for the  
twist angle, electron bulk velocity and thermal map to vary in 
time (but requiring that, for a given epoch, they are the same in the 
different energy bands), and imposing that the geometrical viewing angles 
(see Fig.\ref{fig1}) must be
the same at all epochs. This step provides us with an estimate of
the source parameters and viewing angles. We then use our Monte Carlo code 
to compute the phase averaged spectra, at the various epochs, for the
same sets of parameters and compare with the spectral data. The main 
reason at the basis of this strategy is that we found 
that spectral fitting alone
is unable to constrain the two geometrical angles, while lightcurve 
fitting allows for a better control of complex, time-evolving thermal 
maps.

\begin{figure}
\includegraphics[width=2.in,angle=0]{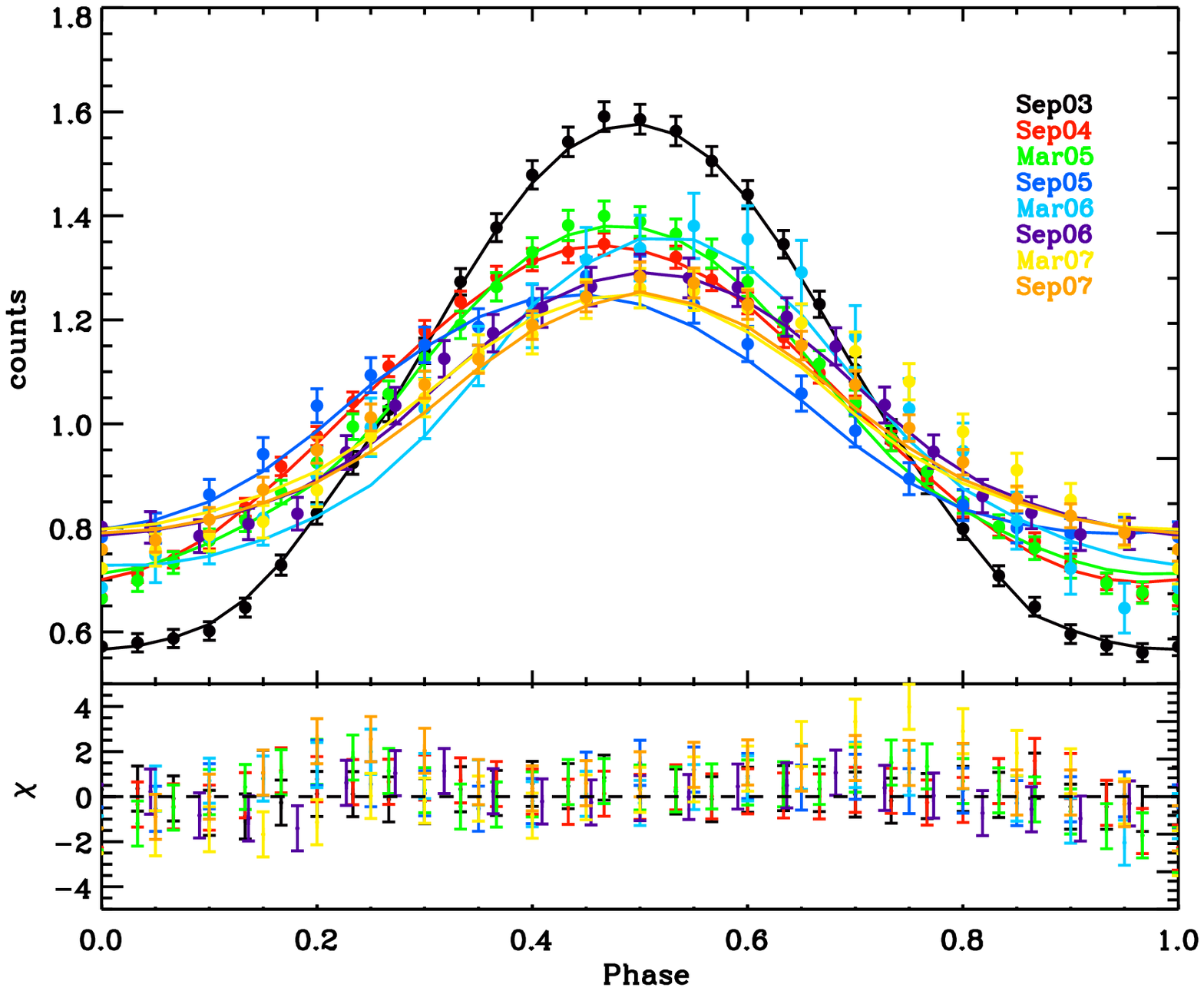}
\includegraphics[width=2.in,angle=0]{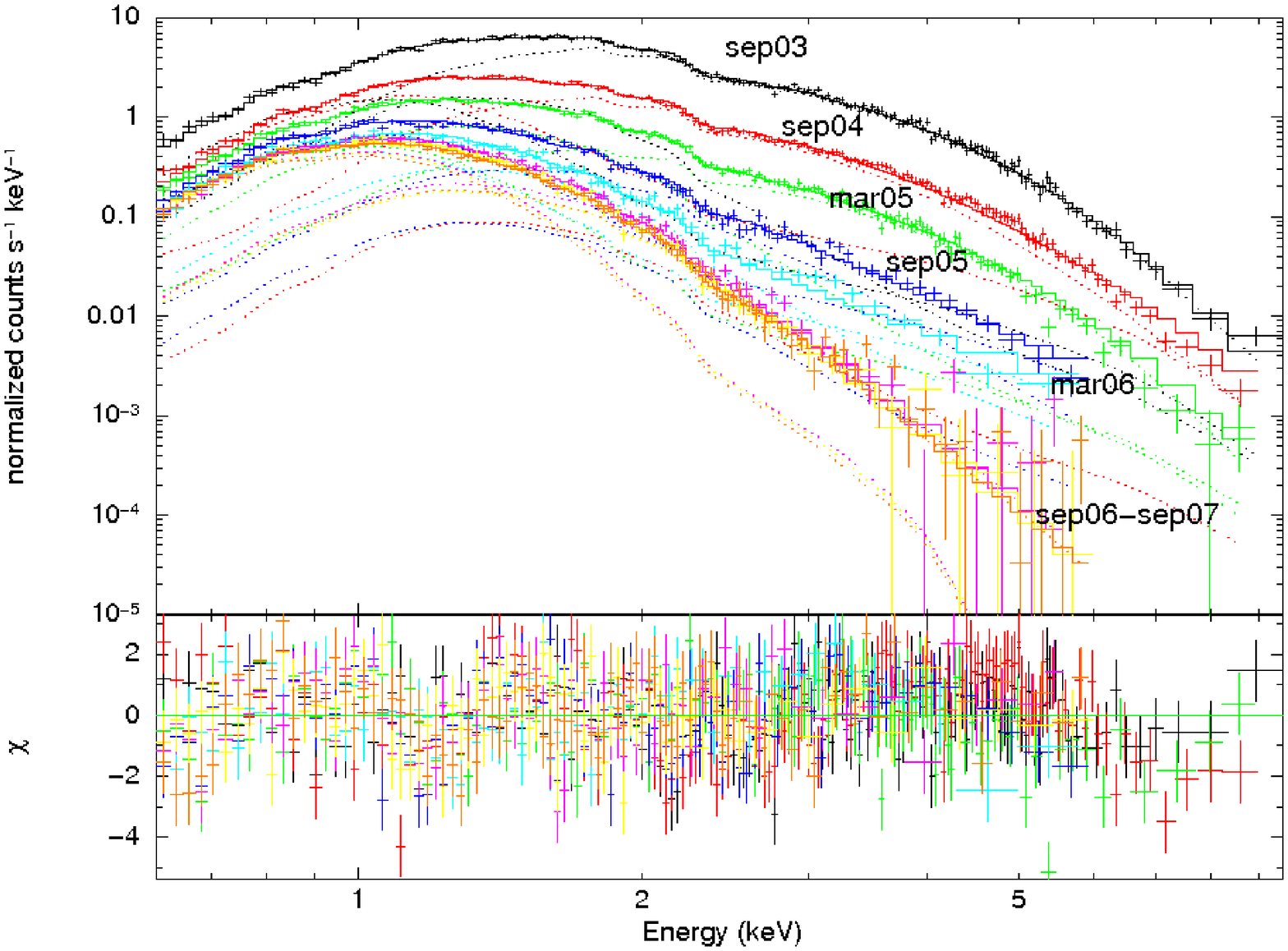}
\caption{Left: 
Synthetic and observed pulse profiles for the eight {\it XMM-Newton} 
observations of XTE J1810-197 in the total energy band. 
Solid lines represent the best-fitting
model, dots the observed lightcurves. Right: 
Spectral evolution in the same observations. Solid lines represent the 
model, while dotted lines refer to its individual components. 
Reproduced by permission of the AAS from
\cite{ale}.\label{fig2}}
\end{figure}

\begin{figure}
\includegraphics[width=2.2in,angle=0]{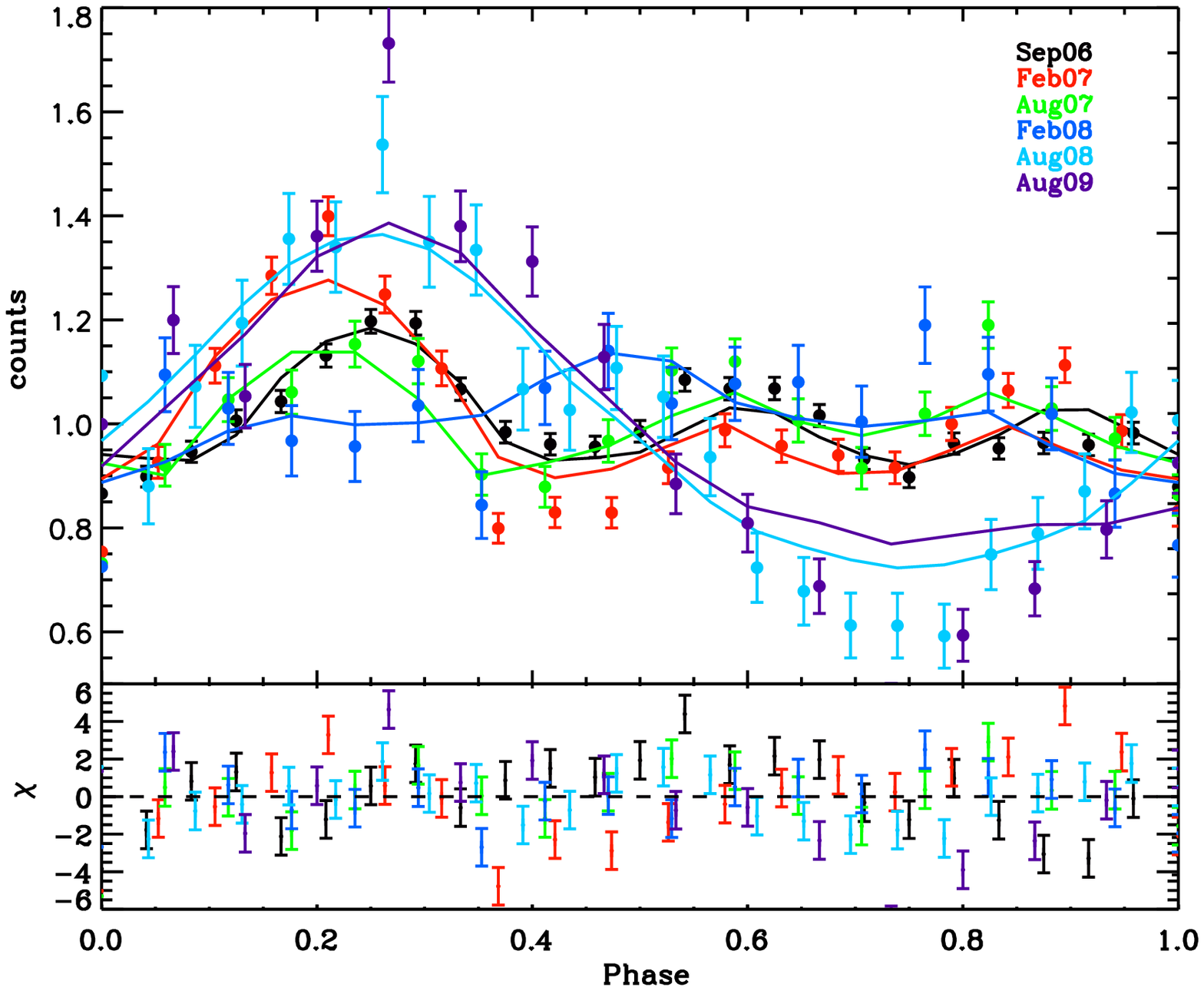}
\includegraphics[height=2.in,angle=0]{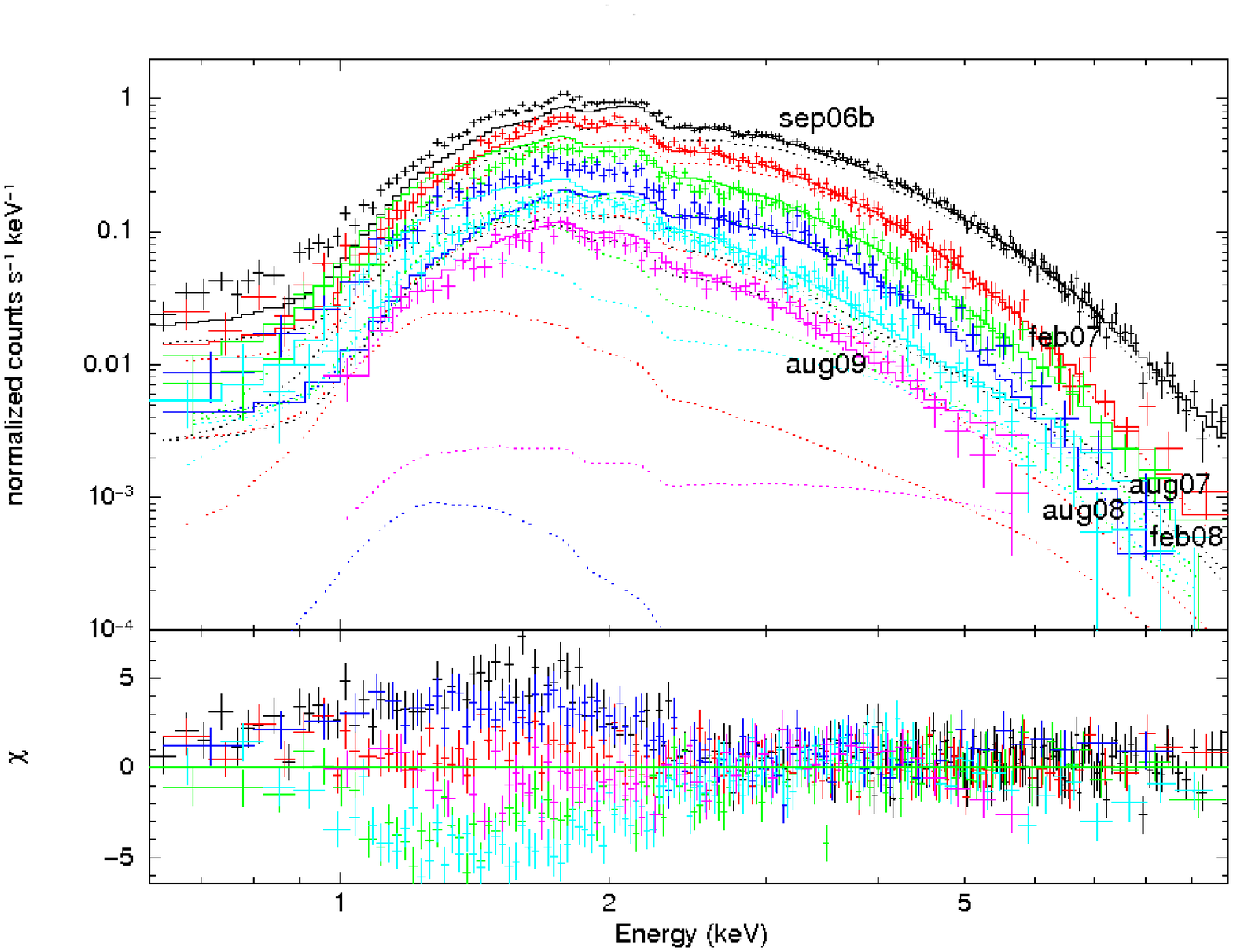}
\caption{Same as in Fig.~\ref{fig2} for CXOU J164710.2-455216. 
Reproduced by permission of the AAS from
\cite{ale}.\label{fig3}}
\end{figure}

For both sources, we find that the best fit is obtained by 
using a three temperature map, made of a hot spot, a warm corona and a 
cooler region (see Fig.~\ref{fig1}). 
In both cases the method was  
successful and we reached a satisfactory fit of both  
the pulse profiles and spectra (see 
Figs.~\ref{fig2},\ref{fig3} and the original paper for details and  
for the values of the fitting parameters).  

\begin{figure}
\includegraphics[width=2.1in,angle=0]{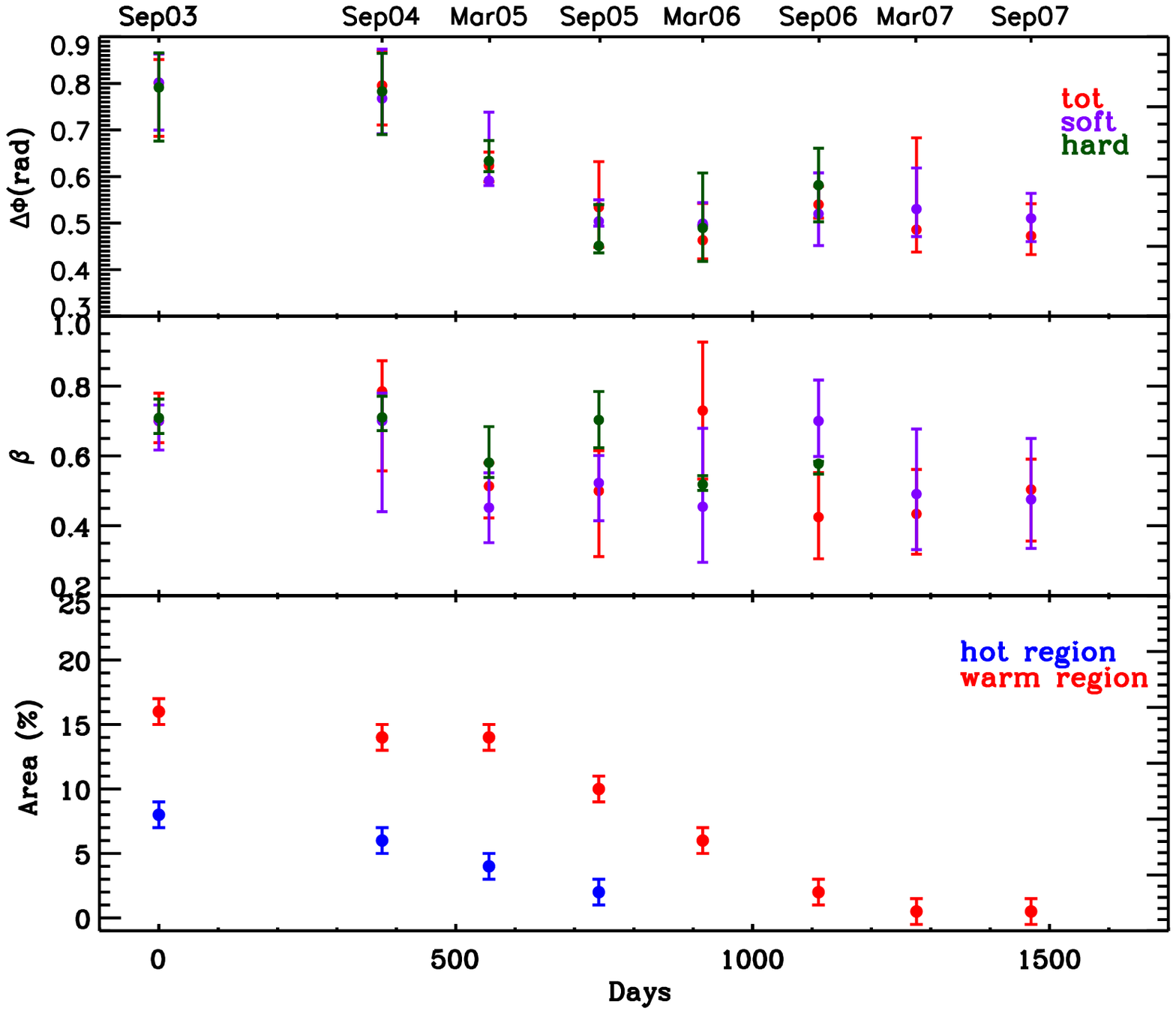}
\includegraphics[width=2.1in,angle=0]{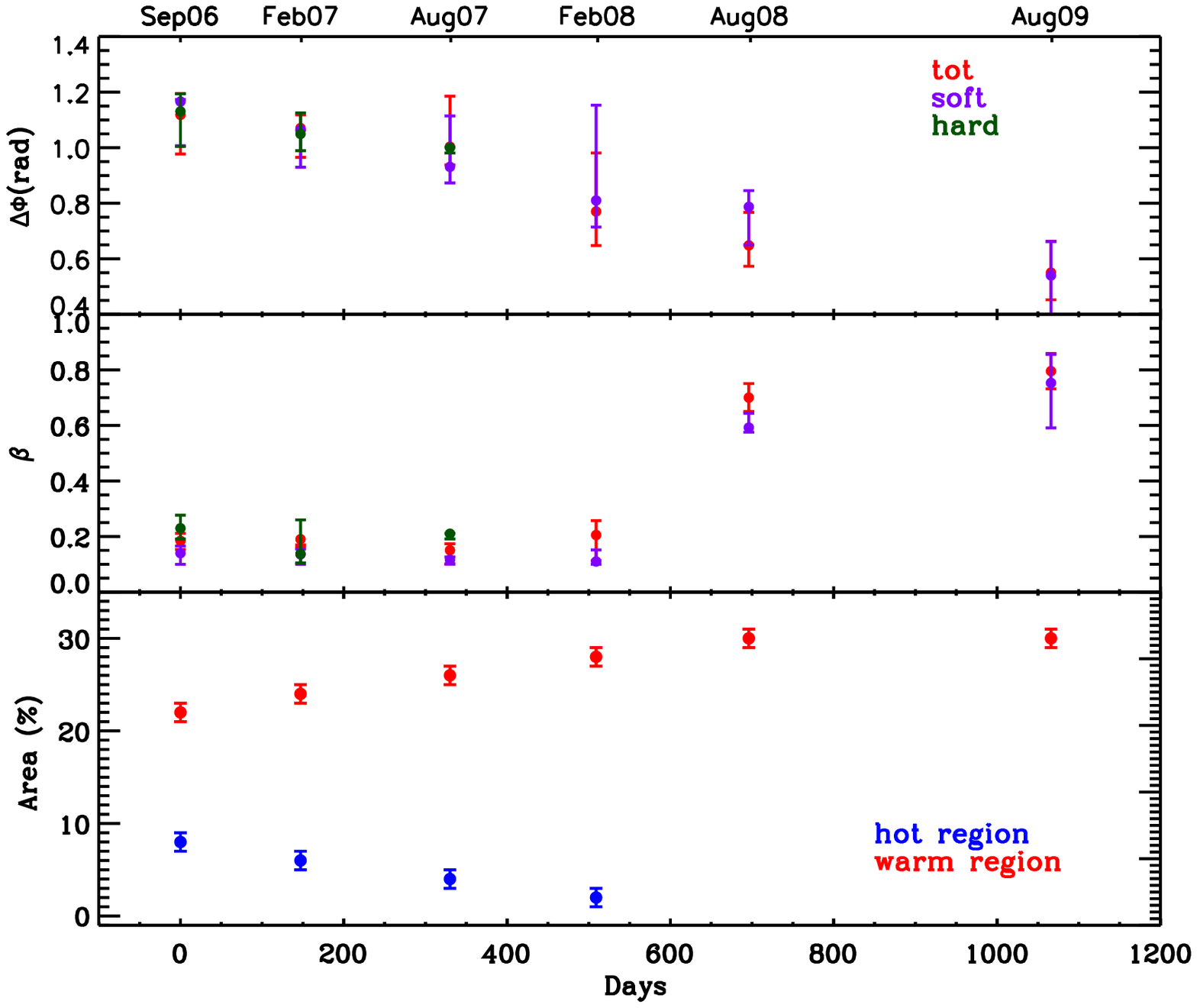}
\caption{Parameters
evolution for XTE J1810-197 (left) and CXOU
J164710.2-455216 (right). From top to bottom: 
twist angle ($\Delta \phi$), bulk velocity ($\beta$) and area of
the warm and hot emitting regions. 
Reproduced by permission of the AAS from
\cite{ale}.
\label{fig4}}
\end{figure}

The evolution of the single spectral components during the outburst decay 
was nevertheless quite different for the two TAXPs. 
For XTE J1810-197 (see Fig.~\ref{fig4}, left), we found that the hot cap 
decreased in size and temperature, until it became indistinguishable from 
the warm corona around March 2006. The warm corona also shrinked, but 
it initially cooled 
down to $\sim 0.3$ keV, and then its temperature remained constant.  It 
was 
still visible in our last observation (Sep 07), 
with a size down to 0.5\% of the NS 
surface. The rest of the NS remained at a temperature comparable with that 
measured in quiescence by  {\it ROSAT} during the entire 
evolution, suggesting that the 
outburst likely involved only a fraction of the star surface \cite[see 
also][]{be09}. The 
best-fitting viewing geometry is  given by $\chi = 148^\circ$, $ \xi= 
23^\circ$. 

The evolution of CXOU J164710.2-455216 was quite different 
(Fig.~\ref{fig4}, right): in this case the hot cap decreased in $T$ and 
size, and the warm corona remained constant at $T \sim 0.45$~keV while it 
increased in size. Interestingly, the area of the hot+warm region 
is constant ($\sim 30\%$ of the NS surface), while the remaining $\sim 
70\%$ is cooler, at $\sim 0.15$~keV. This may suggest that 
the 
thermal map, in quiescence, is made of two components: a warm 
polar region superimposed to the cooler surface. 
The outburst might have heated a portion of the warm cap, producing
the hot zone which then cooled off; we note that the hot spot disappeared 
($\sim$ Aug 08) when the pulse profile switched from three-peaked to 
single-peaked. 
The best-fitting viewing geometry is $\chi = 20^\circ$, $ \xi= 80^\circ$. 

\subsection{Conclusions}

To our knowledge this is the first time that a self-consistent
spectral and timing analysis, based on a realistic modelling of resonant 
scattering, is carried out for magnetars, considering simultaneously a 
large number of datasets over a baseline of years. The results 
support a picture in which only a limited portion of the magnetosphere was 
affected by the outburst (twist), and provide identification for 
the location, meaning and evolution of the various thermal components. 
Future developments will require detailed spectral calculations in a 
magnetosphere with a localized twist which decays in time 
\cite{belo09} and will be matter of future investigations.

\begin{theacknowledgments}
Work partially supported by INAF-ASI through grant AAE I/088/06/0.
\end{theacknowledgments}

\bibliographystyle{aipproc}   

\begin{thebibliography}{9}

\bibitem{ntz}
L. Nobili, R. Turolla, S. Zane, 
\emph{MNRAS} \textbf{386}, 1527-1542 (2008).

\bibitem{za09}
S. Zane, N. Rea, R. Turolla, L. Nobili, 
\emph{MNRAS} \textbf{398}, 1403-1413 (2009).

\bibitem{ale}
A. Albano, R. Turolla, G.L. Israel, S. Zane, L. Nobili,  L. Stella, 
\emph{ApJ} \textbf{722}, 788-802 (2010).

\bibitem{ib04}
A.I. Ibrahim, et al., \emph{ApJ} \textbf{609}, L21-L24 (2004). 

\bibitem{is04}
G.L. Israel, et al., \emph{ApJ} \textbf{603}, L97-L100 (2004).

\bibitem{gh05}
E.V. Gotthelf, \& J.P. Halpern, 
\emph{ApJ} \textbf{632}, 1075-1085 (2005).

\bibitem{gh07}
E.V. Gotthelf, \& J.P. Halpern, 
\emph{Ap\&SS} \textbf{308}, 79-87 (2007).

\bibitem{pe08}
R. Perna, \& E.V. Gotthelf, 
\emph{ApJ} \textbf{681}, 522-529 (2008).

\bibitem{be09}
F. Bernardini, et al., \emph{A\&A}, \textbf{498}, 195-207 (2009).

\bibitem{muno07}
M.P. Muno, et al., \emph{MNRAS}, \textbf{378}, L44-L48 (2007). 

\bibitem{is07}
G.L. Israel, et al., \emph{ApJ}, \textbf{664}, 448-457 (2007).

\bibitem{belo09}
A.M. Beloborodov, \emph{ApJ}, \textbf{703}, 1044-1060 (2009).

\end{thebibliography}

\end{document}